\documentclass[aps,showpacs,epsfig,twocolumn]{revtex4}
\usepackage{epsfig}

\newcommand{\be}{\begin{equation}}
\newcommand{\ee}{\end{equation}}
\newcommand{\bea}{\begin{eqnarray}}
\newcommand{\eea}{\end{eqnarray}}
\usepackage{color}
\begin{document}

\title{\bf\Large {Ferromagnetic phases in spin-Fermion systems}}

\author{Naoum Karchev }

\affiliation{Department of Physics, University of Sofia, 1126 Sofia, Bulgaria }

\begin{abstract}
Spin-Fermion systems which obtain their magnetic properties
from a system of localized magnetic moments being coupled to
conducting electrons are considered. The dynamical degrees of freedom are spin-$s$
operators of localized spins and spin-$1/2$ Fermi operators of
itinerant electrons. Renormalized spin-wave theory, which accounts for the magnon-magnon interaction,
and its extension are developed to describe the two ferrimagnetic phases in
the system: low temperature
phase  $0<T<T^{*}$, where all electrons contribute the ordered ferromagnetic
moment, and high temperature phase $T^{*}<T<T_C$, where only localized spins form
magnetic moment.
The  magnetization as a function of temperature is calculated. The theoretical predictions are utilize
to interpret the experimentally measured magnetization-temperature curves of $UGe_2$..

\end{abstract}

 \maketitle

\section {\bf Introduction}

Spin-Fermion systems, which obtain their magnetic
properties from a system of localized magnetic moments and itinerant electrons are considered.
The true magnons in these systems, which are the transversal fluctuations corresponding
to the total magnetization, are complicated mixtures of the transversal
fluctuations of the spins of localized and itinerant electrons\cite{Karchev08a}. The magnons
interact with localized magnetic moments and itinerant electrons in a different way.
Magnons' fluctuations suppress the ordered moments
of the localized and itinerant electrons at different temperatures. As a result, the
ferromagnetic phase is divided into two phases: low temperature
phase  $0<T<T^{*}$, where all electrons contribute the ordered ferromagnetic
moment, and high temperature phase $T^{*}<T<T_C$, where only localized spins form
magnetic moment.

At first sight the result seems to be counterintuitive because the
moment formed by localized electrons  builds an effective magnetic field, which due to exchange
interaction leads to a finite magnetization of the itinerant electrons.
This is true in the classical limit. In the quantum case the spin
wave fluctuations suppress the magnetic orders at different temperatures $T^*$ and $T_C$ as a
result of a different interaction of magnons with  the localized and itinerant electrons.
The $T^*$ transition is a transition between two magnetically-ordered phases in contrast to the transition
from the magnetically ordered state to the paramagnetic one ($T_C$-transition).

First approach to itinerant electron magnetism, which accounts for the spin
fluctuations, has been developed by Moriya and  Kawabata \cite{MoriyaK,Moriya}.
It is a self-consistent one loop approximation which interpolates between the Heisenberg theory
of localized spins and theory of nearly ferromagnetic metals.
The nonlinear effects of spin fluctuations are treated in
\cite{LonzarichT}, using a self-consistent
rotationally invariant Hartree approximation for itinerant electrons' interaction.

In the spin-Fermion systems the localized spins polarize the itinerant electrons and that feeds back
as an indirect coupling between the localized spins. Averaging in the subspace of the itinerant electrons,
one obtains an effective Heisenberg like
model in terms of the localized spins. This indirect exchange coupling is referred to as
Rudermann-Kittel-Kasuya-Yosida (RKKY) interaction \cite{Kasuya,Yosida,RKittel}.
An approximate but self-consistent theory of spin-Fermion systems is presented in \cite{Nolting97}.
In second-order perturbation theory one gets the well-known RKKY effective interection.
The magnetization curve, the spin polarization of the itinerant electrons and the correlation functions
are worked out in terms of the band occupation and exchange coupling. The subtle point is that the transversal
spin fluctuations are not the true magnon of the system. Therefore the RKKY validity condition requires not only
small Kondo coupling, but it also requires the charge carrier
density to be small, which in turn means that the magnetization of
the mobile electrons is inessential.

In the present paper the Schwinger bosons and slave Fermions are introduced to separate the spin fluctuations
of the electrons from the charge ones. The slave Fermions, which are spinless, are integrated out and
 an effective model in terms of  the transversal fluctuations of the spins of localized and itinerant
electrons is obtained. The anomaly results from the existence of the two separated sources of magnetization.

Renormalized spin-wave (RSW) theory, which accounts for the magnon-magnon interaction,
and its extension are developed to describe the two ferromagnetic phases in the system
and to calculate the magnetization as a function of temperature. It is impossible to require
the theoretically calculated Curie temperature and magnetization-temperature curves to be in
exact accordance with experimental results. The models are idealized, and they do not consider
many important effects. Because of this it is important to formulate theoretical criteria for
adequacy of the method of calculation.
In my opinion the calculations should be in accordance with the Mermin-Wagner theorem \cite{M-W}.
It claims that at nonzero temperature, a one-dimensional or two-dimensional isotropic
spin-S Heisenberg model with finite-range exchange interaction can be neither ferromagnetic
nor antiferromagnetic.
The present method of calculation, being approximate, captures the essentials of the magnon
fluctuations in the theory and satisfy the Mermin-Wagner theorem. The physics of the ferromagnetic
spin-Fermion systems is dominated by the magnon fluctuations and it is important to account for them in the best way.
Comparing figure 4 in the present paper and figure 2 in \cite{Karchev08a} one becomes aware of
the relevance of the present calculations for the accurate reproduction of the basic features of
the system near the characteristic temperatures $T_C$ and $T^*$.

To compare the theoretical results  and the experimental magnetization-temperature curves one has,
first of all, to interpret adequately the measurements. As an example, the experimental
measurements of the ferromagnetic phase of $UGe_2$ are considered . They reveal the presence of an
additional phase line that lies entirely within the ferromagnetic
phase. The characteristic temperature of this transition $T_{x}$,
which is below the Curie temperature $T_C$, decreases with pressure
and disappears at a pressure close to the pressure at which new
phase of coexistence of superconductivity and ferromagnetism
emerges\cite{2fmp1,2fmp2,2fmp3}. The additional phase transition
demonstrates itself through the change in the $T$ dependence of
the ordered ferromagnetic moment\cite{2fmp5,2fmp6,2fmp7}. The
magnetization shows an anomalous enhancement below $T_{x}$.

The paper is organized as follows. In Sec. II an effective model in terms of  the transversal
fluctuations of the spins of localized and itinerant electrons is obtained.
In Sec. III a renormalized spin-wave theory is worked out to calculate the magnetization-temperature curve.
The analysis of the experimental magnetization-temperature curves is given in Sec. IV.
To study the magnetic properties of the $UGe_2$ an effective two magnetic ordered
moments model is considered. Varying the model's parameters, the anomalous temperature dependence
of the magnetization, known from the experiments with $UGe_2$ \cite{2fmp2,2fmp5,2fmp6,2fmp7},
are reproduced theoretically. It is shown that the experimentally measured transition
at temperature $T_x(=T^*)$ is a transition from high temperature phase $T^{*}<T<T_C$,
where only part of the $5f$ uraniun electrons contribute the ordered ferromagnetic
moment, to low temperature phase  $0<T<T^{*}$, where all electrons contribute the magnetization.
A summary in Sec. V concludes the paper.

\section {\bf Effective model}

The dynamical degrees of freedom in spin-Fermion model are spin-$s$ operators of localized
spins and spin-$1/2$ Fermi operators of itinerant electrons.
 One considers a theory with Hamiltonian \bea \label{letter1}\nonumber
h & = & H-\mu N = -t\sum\limits_{  \langle  ij
 \rangle  } {\left( {c_{i\sigma }^ + c_{j\sigma } + h.c.} \right)}
  -\mu \sum\limits_i {n_i
} \\ & - & J^l\sum\limits_{  \langle  ij  \rangle  } {{\bf S}_i
\cdot {\bf S}_j}-J'\sum\limits_{  \langle  ij  \rangle  } {{\bf s}_i
\cdot {\bf s}_j}
  - J\sum\limits_i {{\bf
S}_i}\cdot {\bf s}_i \eea where $s^{\nu}_i=\frac 12
\sum\limits_{\sigma\sigma'}c^+_{i\sigma}
\tau^{\nu}_{\sigma\sigma'}c^{\phantom +}_{i\sigma'}$, with the Pauli
matrices $(\tau^x,\tau^y,\tau^z)$, is the spin of the conduction
electrons, ${\bf S}_i$ is the spin of the localized electrons, $\mu$
is the chemical potential, and $n_i=c^+_{i\sigma}c_{i\sigma}$. The
sums are over all sites of a three-dimensional cubic lattice, and
$\langle i,j\rangle$ denotes the sum over the nearest neighbors. The
Heisenberg terms describe ferromagnetic Heisenberg
exchange between nearest-neighbors localized ($J^l > 0$) and itinerant ($J'>0$) electrons. The last term
in Eq.(\ref{letter1}) describes the ferromagnetic spin-Fermion interaction $(J>0)$.

One represents the Fermi operators in terms of the Schwinger bosons
($\varphi_{i,\sigma}, \varphi_{i,\sigma}^+$) and slave Fermions
($h_i, h_i^+,d_i,d_i^+$)\cite{Karchev08a}. The Bose fields
are doublets $(\sigma=1,2)$ without charge, while Fermions
are spinless with charges 1 ($d_i$) and -1 ($h_i$).
\begin{eqnarray} & & c_{i\uparrow} =
h_i^+\varphi _{i1}+ \varphi_{i2}^+ d_i, \qquad c_{i\downarrow} =
h_i^+ \varphi _{i2}- \varphi_{i1}^+ d_i, \nonumber
\\
& & n_i = 1 - h^+_i h_i +  d^+_i d_i,\quad  s^{\nu}_i=\frac 12
\sum\limits_{\sigma\sigma'} \varphi^+_{i\sigma}
{\tau}^{\nu}_{\sigma\sigma'} \varphi_{i\sigma'},\nonumber
\\& &
\varphi_{i1}^+ \varphi_{i1}+ \varphi_{i2}^+ \varphi_{i2}+ d_i^+
d_i+h_i^+ h_i=1 \label{letter2}
\end{eqnarray}

 Next, we make a change of variables, introducing
Bose doublets $\zeta_{i\sigma}$ and
$\zeta^+_{i\sigma}\,$\cite{2fmp9}
\begin{eqnarray}
\zeta_{i\sigma} & = & \varphi_{i\sigma} \left(1-h^+_i h_i-d^+_i
d_i\right)^
{-\frac 12},\nonumber \\
\zeta^+_{i\sigma} & = & \varphi^+_{i\sigma} \left(1-h^+_i h_i-d^+_i
d_i\right)^ {-\frac 12}, \label{letter6}
\end{eqnarray}
where the new fields satisfy the constraint
$\zeta^+_{i\sigma}\zeta_{i\sigma}\,=\,1$. In terms of the new fields
the spin vectors of the itinerant electrons have the form \be
s^{\nu}_{i}=\frac 12 \sum\limits_{\sigma\sigma'} \zeta^+_{i\sigma}
{\tau}^{\nu}_{\sigma\sigma'} \zeta_{i\sigma'} \left[1-h^+_i
h_i-d^+_i d_i\right], \label{letter8} \ee
where the unit vector $\,\,
n^{\nu}_i=\sum\limits_{\sigma\sigma'} \zeta^+_{i\sigma}
{\tau}^{\nu}_{\sigma\sigma'} \zeta_{i\sigma'}\,\,\, ({\bf
n}_i^2=1)\,\,$ identifies the local orientation of the spin of the
itinerant electron\cite{Karchev08a}. Let us average the spin of electrons in the
subspace of the Fermions $(d^+_i, \,d_i)$ and $(h^+_i,\,h_i)$ (to
integrate the Fermions out in the path integral approach). One obtains
\begin{eqnarray}\label{letter9}
{\bf s}_{i} & = & m {\bf n}_i \qquad  {\bf s}_{i}^2=m^2  \\
m & = & \frac 12 \left(1-<h^+_i h_i>_f-<d^+_id_i>_f\right),\nonumber
\end{eqnarray}
where $<.....>_f$  means an average in the subspace of the Fermions $d(d^+)$ and $h(h^+)$
when the spin fluctuations of the itinerant electrons are set equal to zero.
Hence, the amplitude of the spin vector $"m"$ is an effective spin of
the itinerant electrons accounting for the fact that some sites, in
the ground state, are doubly occupied or empty.

It is more convenient to use the rescaled Bose fields
\begin{equation}
\xi_{i\sigma}=\sqrt{2m}\,\zeta_{i\sigma},\qquad\qquad
\xi^+_{i\sigma}=\sqrt{2m}\,\zeta^+_{i\sigma} \label{letter10}
\end{equation}
which satisfy the constraint $\xi^+_{i\sigma}\xi_{i\sigma}=2m$, and to
introduce the vector,
\begin{equation}
 M^{\nu}_{i}=\frac 12 \sum\limits_{\sigma\sigma'} \xi^+_{i\sigma}
{\tau}^{\nu}_{\sigma\sigma'} \xi_{i\sigma'}\quad {\bf M}_{i}^2=m^2 .
\label{letter12}
\end{equation}
Then, the spin-vector of itinerant electrons can be written in the form
\begin{equation}
{\bf s}_{i}=\frac {1}{2m}{\bf M}_{i}\left(1-h^+_i\,h_i\,-\,
d^+_i\,d_i\right) \label{letter13}
\end{equation}
and ${\bf M_i}=<{\bf s}_{i}>_f$

 The Hamiltonian is quadratic with
respect to the Fermions $d_i, d^+_i$ and $h_i, h^+_i$, and one can
average in the subspace of these Fermions (to integrate them out in
the path integral approach). As a result, we obtain an effective
theory of two vectors ${\bf S}_i$ and ${\bf M}_i$ with
Hamiltonian
\begin{equation}
 h_{eff}= -  J^l\sum\limits_{  \langle  ij  \rangle  } {{\bf S}_i \cdot
{\bf S}_j}-  J^{it}\sum\limits_{  \langle  ij  \rangle  } {{\bf M}_i
\cdot {\bf M}_j}
  - J\sum\limits_i {{\bf
S}_i}\cdot {\bf M}_i \label{letter14}
\end{equation}
The first term is the term which describes the exchange of localized
spins in the Hamiltonian Eq.(\ref{letter1}). The second term has two components:
one is the term in the Hamiltonian Eq.(\ref{letter1})
which describes the exchange of the spins of the itinerant electrons, while
the second one is obtained integrating out the Fermions.
It is calculated in the one
loop approximation and in the limit when the frequency and the wave
vector are small. For the effective exchange constant $J^{it}$, at
zero temperature, one obtains
\bea\label{letter14a} & & J^{it} = J'\\
& + & \frac {t}{6m^2}\frac 1N
\sum\limits_{k}\left(\sum\limits_{\nu=1}^3\cos
k_{\nu}\right)\left[\theta(-\varepsilon^d_k)-\theta(-\varepsilon^h_k)\right]
\nonumber \\
& - & \frac {2t^2}{3m^2 s J}\frac 1N
\sum\limits_{k}\left(\sum\limits_{\nu=1}^3\sin^2
k_{\nu}\right)\left[1-\theta(-\varepsilon^h_k)-\theta(-\varepsilon^d_k)\right]\nonumber\eea
where $N$ is the number of lattice's sites, $\varepsilon^h_k$ and
$\varepsilon^d_k$  are Fermions' dispersions, \bea\label{letter14b}
\varepsilon^h_k & = & 2t(\cos k_x+\cos k_y+\cos k_z)+s J/2 +\mu  \\
\varepsilon^d_k & = & -2t(\cos k_x+\cos k_y+\cos k_z)+s J/2 -\mu,
\nonumber \eea
 and wave vector $k$ runs over the first
Brillouin zone of a cubic lattice. Calculating the ratio $ J^{it}/J$ from the equation (\ref{letter14a})
one obtains that the second term,
which comes from the tadpole diagram with one $d$ or $h$ line, is proportional to $t/J$ and the last term,
which results from the calculation of loop diagrams with two $d$ or $h$ lines is proportional to $(t/J)^2$.
This means that our one loop approximation is most relevant for small $t/J$.
 For the case of experimental interest the density of
itinerant electrons per lattice site is equal to one and then the contribution of the spin-Fermion
interaction to the exchange constant $J^{it}$ Eq.(\ref{letter14a}) is negative.
As a result $J^{it}$ is positive but very small compare with $J^{l}$ and $J$.
The third term in Eq.(\ref{letter14}) is obtained from the last one in the
Hamiltonian Eq.(\ref{letter1}) using the representation Eq.(\ref{letter13}) for
the spin of itinerant electrons and Eq.(\ref{letter9}).

\section {\bf Renormalized spin-wave theory}

We are going to study the ferromagnetic phase of the two-spin system
Eq.(\ref{letter14}) with $J^l>0,\,\,J^{it}>0$, and $J>0$. To proceed we
use the Holstein-Primakoff representation of the spin vectors ${\bf
S}_j(a^+_j,a_j)$ and ${\bf M}_j(b^+_j,\,b_j)$
\bea\label{rsw2} & &
S_{j}^+ = S^1_{j} + i S^2_{j}=\sqrt {2s - a^+_ja_j}\,\,\,\,a_j \nonumber \\
& & S_{j}^- = S^1_{j} - i S^2_{j}=a^+_j\,\,\sqrt {2s-a^+_ja_j} \nonumber
\\ & & S^3_{j} = s - a^+_ja_j  \\
& &
M_{j}^+ = M^1_{j} + i M^2_{j}=\sqrt {2m-b^+_jb_j}\,\,\,\,b_j \nonumber \\
& & M_{j}^- = M^1_{j} - i M^2_{j}=b^+_j\,\,\sqrt {2m-b^+_jb_j} \nonumber
\\ & & M^3_{j} = m - b^+_jb_j \nonumber \eea
where $a^+_j,\,a_j$
and $b^+_j,\,b_j$ are Bose fields, while $s$ and $m$ are the effective spins of the localized and itinerant electrons.
In terms of the Bose fields and keeping only the quadratic and quartic
terms, the effective Hamiltonian Eq.(\ref{letter14}) adopts the form
\be\label{heff}
h_{eff}\,=\,h_2\,+\,h_4
\ee
where
\bea\label{letter17}
 h_2& = & s\,J^l\sum\limits_{  \langle  ij  \rangle
 }(a_i^+a_i+a_j^+a_j-a_j^+a_i-a_i^+a_j)\nonumber\\
 & + & m\,J^{it} \sum\limits_{  \langle  ij  \rangle
 }(b^+_ib_i+b^+_jb_j-b^+_jb_i-b^+_ib_j)\\
 & - & J\sum\limits_i
 (\sqrt{sm}\,[a_i^+b_i+b_i^+a_i]-sb_i^+b_i-ma_i^+a_i)\nonumber
 \eea
 \bea\label{heff2}
 h_4 & = & \frac {J^l}{4} \sum\limits_{  \langle  ij  \rangle
 }[a_i^+a_j^+(a_i-a_j)^2\,+\,(a_i^+-a_j^+)^2a_i a_j]\nonumber\\
 & + & \frac {J^{it}}{4} \sum\limits_{  \langle  ij  \rangle
 }[b_i^+b_j^+(b_i-b_j)^2\,+\,(b_i^+-b_j^+)^2b_i b_j]\nonumber\\
 & + & \frac {J}{4} \sum\limits_i
 [\sqrt{\frac {s}{m}}(a_i^+b_i^+b_ib_i\,+\,b_i^+b_i^+b_ia_i) \\
 & + &  \sqrt {\frac {m}{s}}( b_i^+a_i^+a_ia_i\,+\,a_i^+a_i^+a_ib_i)\,-\,4 a_i^+a_ib_i^+b_i]\nonumber
 \eea
 and terms without fields are dropped.

 The next step is to represent the Hamiltonian in the Hartree-Fock  approximation:
 \be\label{HF1}
 h_{eff}\approx h_{HF}=h_{cl}+h_q\ee where
 \bea\label{HF2}
 h_{cl}& = & 3 N J^l s^2 (u^l-1)^2+ 3 N J^{it} m^2 (u^{it}-1)^2 \nonumber \\
& + &  N J s m (u-1)^2,
\eea
\bea\label{HF3}
 h_q & = & s\,J^l u^l\sum\limits_{  \langle  ij  \rangle
 }(a_i^+a_i+a_j^+a_j-a_j^+a_i-a_i^+a_j)\nonumber\\
 & + & m\,J^{it} u^{it} \sum\limits_{  \langle  ij  \rangle
 }(b^+_ib_i+b^+_jb_j-b^+_jb_i-b^+_ib_j)\\
 & - & J u \sum\limits_i
 (\sqrt{sm}\,[a_i^+b_i+b_i^+a_i]-sb_i^+b_i-ma_i^+a_i)\nonumber
 \eea
 Equation (\ref{HF3}) shows that the Hartree-Fock parameters $u^l,\,u^{it}$ and $u$ renormalize
the exchange constants $J^l,\,J^{it}$ and $J$, respectively.

 It is convenient to rewrite the Hamiltonian in the momentum space representation:
 \be \label{letter18}
 h_{q} = \sum\limits_{k}\left (\varepsilon^a_k\,a_k^+a_k\,+\,\varepsilon^b_k\,b_k^+b_k\,-
 \,\gamma\,(a_k^+b_k+b_k^+a_k)\,\right),\ee
 where the wave vector $k$ runs over the first Brillouin zone $B$ of a cubic lattice.
 The dispersions are given by the equalities
 \bea\label{dispersion1}
\varepsilon^a_k & = & 2s\,J^l\,u^l\varepsilon_k
\,+\,m\,J u \nonumber\\
\varepsilon^b_k & = & 2m \,J^{it}\,u^{it} \varepsilon_k \,+\,s\,J\,u \\
\gamma & = & J\,u\,\sqrt{s\,m}\nonumber\eea
with
\be\label{dispersion2}
\varepsilon_k=3-\cos k_x-\cos k_y-\cos k_z.
\ee

To diagonalize the Hamiltonian, one introduces new Bose fields
$\alpha_k,\,\alpha_k^+,\,\beta_k,\,\beta_k^+$,
\bea\label{letter21}
a_k & = & \cos\theta_k\,\alpha_k\,+\,\sin\theta_k\,\beta_k,  \nonumber \\
\\
b_k & = & -\sin\theta_k\,\alpha_k\,+\,\cos\theta_k\,\beta_k \nonumber
\eea
with coefficients of transformation, \bea\label{letter22}
\cos\theta_k & = & \sqrt{\frac 12\,\left (1+\frac
{\varepsilon^a_k-\varepsilon^b_k}{\sqrt{(\varepsilon^a_k-\varepsilon^b_k)^2+4\gamma^2}}\right
)},\nonumber \\
\\
\sin\theta_k & = & \sqrt{\frac 12\,\left (1-\frac
{\varepsilon^a_k-\varepsilon^b_k}{\sqrt{(\varepsilon^a_k-\varepsilon^b_k)^2+4\gamma^2}}\right
)}\nonumber
\eea
The transformed
Hamiltonian adopts the form \be \label{letter23} h_{q} = \sum\limits_{k}\left
(E^{\alpha}_k\,\alpha_k^+\alpha_k\,+\,E^{\beta}_k\,\beta_k^+\beta_k\right),
\ee with new dispersions \bea  \label{letter24}
E^{\alpha}_k  & = & \frac 12\,\left
[\varepsilon^a_k+\varepsilon^b_k\,+ \,
\sqrt{(\varepsilon^a_k-\varepsilon^b_k)^2+4\gamma^2}\right] \nonumber \\
\\
E^{\beta}_k  & = & \frac 12\,\left
[\varepsilon^a_k+\varepsilon^b_k\,- \,
\sqrt{(\varepsilon^a_k-\varepsilon^b_k)^2+4\gamma^2}\right] \nonumber \eea

With positive
exchange constants $J^l,\,J^{it},J$ and positive Hartree-Fock parameters $u^l,\,u^{it},\,u$ the Bose fields'
dispersions are positive $\varepsilon^a_k>0,\,\,\varepsilon^b_k>0$
for all values of $k\in B$. As a result, $E^{\alpha}_k>0$ and
$E^{\beta}_k\geq 0$ with $E^{\beta}_0=0$. Near the zero wave vector,
$E^{\beta}_k\approx \rho k^2$ where the spin-stiffness constant is
\be\label{stiffness} \rho=\frac {(s^2 J^l u^l\,+\,m^2 J^{it}u^{it})}{(s+m)}.\ee Hence, $\beta_k$ is the
long-range \textbf{(magnon)} excitation in the two-spin effective
theory, while $\alpha_k$ is a gapped excitation with gap
$E^{\alpha}_0\,=\,(s+m)J u$.

To obtain the system of equations for the Hartree-Fock parameters we consider
the free energy of a system with Hamiltonian $h_{HF}$ equations (\ref{HF1}), (\ref{HF2}) and (\ref{letter23}):
\bea\label{2FMPhaseFreeE}
\mathcal{F}&=& 3J^l s^2 (u^l-1)^2+ 3J^{it} m^2 (u^{it}-1)^2
 + J s m (u-1)^2 \nonumber \\
 \\
& + & \frac {1}{\beta N} \sum\limits_{k}\left[ \ln\left(1-e^{-\beta E^{\alpha}_k}\right)\,+\,\ln\left(1-e^{-\beta E^{\beta}_k}\right)\right],\nonumber\eea
where $\beta\,=\,1/T$\,\, is the inverse temperature. Then the three equations for the Hartree-Fock parameters
\be\label{rsw13}\partial\mathcal{F}/\partial u^l=0,\quad \partial\mathcal{F}/\partial u^{it}=0,\quad\partial\mathcal{F}/\partial u=0\ee
have the form (see the appendix)
\bea\label{rsw14} u^l & = & 1-\frac {1}{3s} \frac 1N \sum\limits_{k} \varepsilon_k \left[\cos^2\theta_k \,n_k^{\alpha}\, +\, \sin^2 \theta_k\, n_k^{\beta}\right]\nonumber \\
u^{it} & = & 1-\frac {1}{3 m} \frac 1N \sum\limits_{k} \varepsilon_k \left[\sin^2\theta_k \,n_k^{\alpha}\, +\, \cos^2\theta_k\, n_k^{\beta}\right]\nonumber \\
u & = & 1-\frac 1N \sum\limits_{k}\left[\left(\frac {1}{2s}\cos^2\theta_k+\frac {1}{2m}\sin^2\theta_k \right)\,n_k^{\alpha}\right.\nonumber \\
& + & \left.\left(\frac {1}{2m}\cos^2\theta_k+\frac {1}{2s}\sin^2\theta_k \right) \,n_k^{\beta} \right. \\
& + & \left. \frac {J u}{\sqrt{(\varepsilon^a_k\,-\,\varepsilon^b_k)^2\,+\,4\gamma^2}} \,(n_k^{\alpha}-n_k^{\beta})\right]\nonumber
\eea
where $n_k^{\alpha}$ and $n_k^{\beta}$ are the Bose functions of $\alpha_k$ and $\beta_k$ excitations. The Hartree-Fock parameters, the solution of the system of equations (\ref{rsw14}), are positive functions of $T/J$, $u^l(T/J)>0,\,u^{it}(T/J)>0$ and $u(T/J)>0$. Utilizing these functions, one can calculate the spontaneous magnetization of the system, which is a sum of the spontaneous magnetization of the localized and itinerant electrons $M\,=\,M^l\,+\,M^{it}$. In terms of the Bose functions of the $\alpha_k$ and $\beta_k$ excitations they adopt the form
\bea\label{letter26}& & M^l\,=\,s-\frac 1N \sum\limits_{k}\left
[\cos^2\theta_k\,n^{\alpha}_k\,+\,\sin^2\theta_k\,n^{\beta}_k
\right], \nonumber
\\
& & M^{it}\,=\,m-\frac 1N \sum\limits_{k}\left
[\sin^2\theta_k\,n^{\alpha}_k\,+\,\cos^2\theta_k\,n^{\beta}_k\right],\nonumber \\
& & M\,=\,s\,+\,m-\frac 1N \sum\limits_{k}\left
[n^{\alpha}_k\,+\,n^{\beta}_k\right].\eea
The magnetization depends on the
dimensionless temperature $T/J$ and dimensionless parameters
$s,\,m,\,J^l/J$ and $J^{it}/J$. For parameters $s=1,\,m=0.3,\,J^l/J=0.25$ and $J^{it}/J=0.0025$ the
functions $M^l(T/J)$ and $M^{it}(T/J)$ are depicted in
figure 1.  The upper (black) line is the magnetization of the localized electrons $M^l$, the bottom (red) line is the magnetization of the itinerant electrons $M^{it}$.
\begin{center}
\begin{figure}[htb]
\label{2FMPhIIfig1}
\centerline{\psfig{file=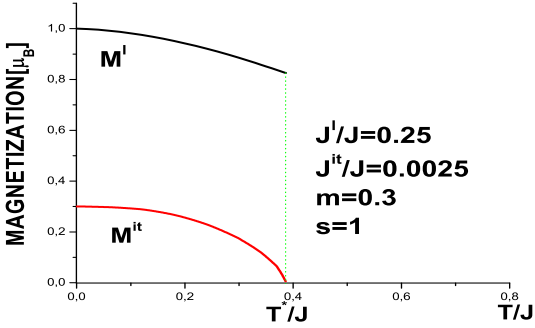,width=8cm,height=5cm}}
\caption{(color online)\,Temperature dependence of the spontaneous magnetization for parameters
$s=1,\,m=0.3,\,J^l/J=0.25$ and $J^{it}/J=0.0025$:
$M^l$ (black line)-magnetization of the localized electrons,
$M^{it}$ (red line)-magnetization of the itinerant electrons.
$T^*$ is the temperature at which the magnetization of the itinerant electrons becomes equal to zero}
\end{figure}
\end{center}

At characteristic temperature $T^*$  spontaneous magnetization of itinerant electrons becomes equal to zero, while spontaneous magnetization of localized spins is still nonzero.
This is because the magnon excitation $\beta_k$ in the effective theory
Eq.(\ref{letter14}) is a complicated mixture of the transversal
fluctuations of the spins of localized and itinerant electrons
Eq.(\ref{letter21}). As a result, the magnons' fluctuations suppress
in a different way the magnetic order of these electrons.
Above $T^*$ the system of equations (\ref{rsw14}) has no solution and one has to modify the renormalized spin-wave theory.

\subsection{\bf Modified RSW theory}

To formulate mathematically the modified RSW theory one introduces \cite{Karchev08a}
two parameters $\lambda^l$ and $\lambda^{it}$ to enforce the magnetic
moments both of the localized and the itinerant electrons to be
equal to zero in paramagnetic phase. To this end, we add two new terms
to the effective Hamiltonian Eq.(\ref{letter14}), \be
\label{letter28} \hat{h}_{eff}\,=\,h_{eff}\,-\,\sum\limits_i \left
[\lambda^l S^z_i\,+\,\lambda^{it} M_i^z\right]. \ee
In Hartree-Fock approximation, in momentum
space, the Hamiltonian adopts the form
\be \label{letter18}
 \hat {h}_{q} = \sum\limits_{k}\left (\hat{\varepsilon}^a_k\,a_k^+a_k\,+\,\hat{\varepsilon}^b_k\,b_k^+b_k\,-
 \,\gamma\,(a_k^+b_k+b_k^+a_k)\,\right),\ee
 where the the new dispersions are
 \be \label{dispersion2a}
\hat{\varepsilon}^a_k\,=\varepsilon^a_k\,+\,\lambda^l, \qquad
\hat{\varepsilon}^b_k\,=\varepsilon^b_k\,+\,\lambda^{it}.\ee

Utilizing the same transformation Eq.(\ref{letter21}) with coefficients
\bea\label{letter22b}
\cos\hat{\theta}_k & = & \sqrt{\frac 12\,\left (1+\frac
{\hat{\varepsilon}^a_k-\hat{\varepsilon}^b_k}{\sqrt{(\hat{\varepsilon}^a_k-\hat{\varepsilon}^b_k)^2+4\gamma^2}}\right
)},\nonumber \\
\\
\sin\hat{\theta}_k & = & \sqrt{\frac 12\,\left (1-\frac
{\hat{\varepsilon}^a_k-\hat{\varepsilon}^b_k}{\sqrt{(\hat{\varepsilon}^a_k-\hat{\varepsilon}^b_k)^2+4\gamma^2}}\right
)}\nonumber
\eea
one obtains the Hamiltonian in diagonal form
\be \label{letter23b} \hat{h}_{q} = \sum\limits_{k}\left
(\hat{E}^{\alpha}_k\,\alpha_k^+\alpha_k\,+\,\hat{E}^{\beta}_k\,\beta_k^+\beta_k\right),
\ee
where
\bea  \label{letter24b}
\hat{E}^{\alpha}_k  & = & \frac 12\,\left
[\hat{\varepsilon}^a_k+\hat{\varepsilon}^b_k\,+ \,
\sqrt{(\hat{\varepsilon}^a_k-\hat{\varepsilon}^b_k)^2+4\gamma^2}\right] \nonumber \\
\\
\hat{E}^{\beta}_k  & = & \frac 12\,\left
[\hat{\varepsilon}^a_k+\hat{\varepsilon}^b_k\,- \,
\sqrt{(\hat{\varepsilon}^a_k-\hat{\varepsilon}^b_k)^2+4\gamma^2}\right]. \nonumber \eea

It is convenient to represent the
parameters $\lambda^l$ and $\lambda^{it}$ in the form
\be\label{lambdamu}
\lambda^l\,=\,m J (\mu^l\,-\,1), \qquad \lambda^{it}\,=\,s J
(\mu^{it}\,-\,1).\ee
In terms of the parameters $\mu^l$ and
$\mu^{it}$, the dispersions adopt the form
\bea\label{dispersionmu}
\hat{\varepsilon}^a_k & = & 2sJ^l\,u^l\varepsilon_k+mJ u\mu^l \nonumber \\
\hat{\varepsilon}^b_k & = & 2mJ^{it}u^{it}\varepsilon_k+sJu\mu^{it} \eea
The renormalized spin-wave theory is reproduced when
$\mu^l=\mu^{it}=1$($\lambda^l=\lambda^{it}=0$). We assume $\mu^l$
and $\mu^{it}$ to be positive ($\mu^l>0,\,\mu^{it}>0$). Then,
$\hat{\varepsilon}^a_k>0$, $\hat{\varepsilon}^b_k>0$, and
$\hat{E}^{\alpha}_k>0$ for all values of the wave-vector $k$.
The $\beta_k$ dispersion is
non-negative, $\hat{E}^{\beta}_k\geq0$ if $\mu^l \mu^{it}\geq1$. In
the particular case $\mu^l \mu^{it}=1$\,\, $\hat{E}^{\beta}_0=0$,
and, near the zero wave vector, $\hat{E}^{\beta}_k\approx \hat{\rho}
k^2$ with spin-stiffness constant equals
\be\label{rhohat}
\hat{\rho}=\frac {s^2 J^l u^l
\mu^{it}+m^2 J^{it}u^{it} \mu^l}{s \mu^{it}+m \mu^l}.\ee Hence, in this
case, $\beta_k$ boson is the long-range excitation (magnon) in the
system. In the case $\mu^l \mu^{it}>1$, both $\alpha_k$ boson and
$\beta_k$ boson are gapped excitations.

The parameters $\lambda^l$ and $\lambda^{it}$ ($\mu^l, \mu^{it}$) are introduced
to enforce the spontaneous magnetizations of the localized and itinerant electrons
to be equal to zero in the paramagnetic phase. One finds out the parameters
$\mu^l$ and $\mu^{it}$, as well as the Hartree-Fock parameters, as functions of temperature,
solving the system of five equations, equations (\ref{rsw14}) and the equations $M^l=M^{it}=0$,
where the  spontaneous magnetizations have the same representation as equations (\ref{letter26})
but with coefficients $\cos\hat{\theta}_k,\,\,
\sin\hat{\theta}_k$, and dispersions $\hat{E}^{\alpha}_k,\,\,
\hat{E}^{\beta}_k$ in the expressions for the Bose functions.
The numerical calculations show that for high enough temperature
$\mu^l\mu^{it}>1$. When the temperature decreases the product $\mu^l\mu^{it}$ decreases,
remaining larger than one. The temperature at which the product becomes equal
to one ($\mu^l\mu^{it}=1$) is the Curie temperature.

Below $T_C$, the spectrum contains magnon
excitations, thereupon $\mu^l\mu^{it}=1$. It is convenient to
represent the parameters in the following way:
\be\label{letter39}\mu^{it}=\mu, \quad\quad \mu^l=1/\mu.\ee

In the ordered phase magnon excitations are the origin of the suppression of the magnetization.
Near the zero temperature their contribution is small and at zero temperature
spontaneous magnetizations $M^l$ and $M^{it}$ reach their saturations $(M^l=s, \,\,M^{it}=m)$.
On increasing the temperature magnon fluctuations suppress the magnetization of localized and
itinerant electrons in different ways.
At $T^*$ the magnetization of the itinerant electrons $M^{it}$ becomes equal to zero.
Increasing the temperature above $T^*$, $M^{it}$ should be zero. This is why we impose the condition
$M^{it}(T)=0$ if $T>T^{*}$. For temperatures above $T^*$, the parameter $\mu$ and
the Hartree-Fock parameters are solution of a system of four equations, equations (\ref{rsw14})
with $\cos\hat{\theta}_k,\,\,\sin\hat{\theta}_k,\,\,\hat{\varepsilon}_k^a,\,\,\hat{\varepsilon}_k^b,\,\,\hat{E}^{\alpha}_k,\,\,\hat{E}^{\beta}_k$ instead of $\cos\theta_k,\,\,\sin\theta_k,\,\,\varepsilon_k^a,\,\,\varepsilon_k^b,\,\,E^{\alpha}_k,\,\,E^{\beta}_k$,
and the equation $M^{it}=0$. The Hartree-Fock parameters, as a functions of temperature $T/J$,
are depicted in figure 2 for parameters $s=1,\,m=0.3,\,J^l/J=0.25$ and $J^{it}/J=0.0025$.
The vertical dotted (green) line corresponds to $T^*/J$.
The function $\mu(T/J)$ is depicted in figure 3 for the same parameters.

One utilizes the
obtained functions $\mu(T)$, $u^l(T)$, $u^{it}(T)$, $u(T)$ to calculate the spontaneous magnetization
as a function of the temperature. Above $T^*$, the magnetization of
the system is equal to the magnetization of the localized electrons.
For the same parameters as above the functions $M^l(T/J)$ and $M^{it}(T/J)$ and
$M(T/J)\,=\,M^l(T/J)\,+\,M^{it}(T/J)$  are depicted in figure 4.
The upper (black) line is the magnetization of localized electrons $M^l(T/J)$,
the middle (red) line is the magnetization of the itinerant electrons  $M^{it}(T/J)$
and the bottom (blue) line is the total magnetization $M(T/J)$. Comparing figure 4, in the present paper,
and figure 2 in \cite{Karchev08a} one becomes aware of the relevance of the present
calculations for the accurate reproduction of the basic features of
the system near the characteristic temperatures $T_C$ and $T^*$.
\begin{center}
\begin{figure}[htb]
\label{letterfig2}
\centerline{\psfig{file=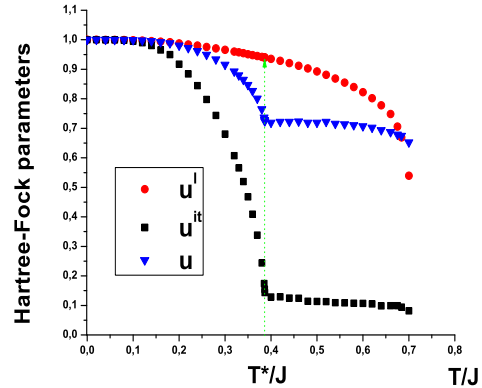,width=7cm,height=6cm}}
\caption{(color online)\, Hartree-Fock parameters $u^l$,\,$u^{it}$ and $u$ as a function of $T/J$ for $s\,=\,1,\,m\,=\,0.3,\,J^l/J\,=\,0.25$ and $J^{it}/J\,=\,0.0025$. The vertical dotted (green) line corresponds to $T^*/J$  }\label{fig2}
\end{figure}
\end{center}
\begin{center}
\begin{figure}[htb]
\label{letterfig5}
\centerline{\psfig{file=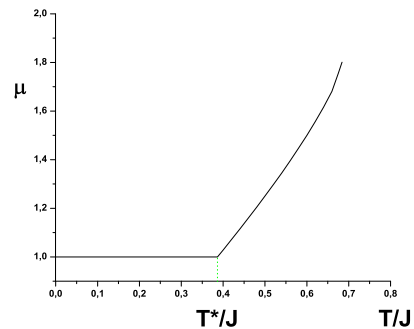,width=6cm,height=5cm}}
\caption{(color online)\,\,$\mu(T/J)$ for parameters $s\,=\,1,\,m\,=\,0.3,\,J^l/J\,=\,0.25$ and $J^{it}/J\,=\,0.0025$.
The vertical dotted (green) line corresponds to $T^*/J$ }
\end{figure}
\end{center}
\begin{center}
\begin{figure}[htb]
\label{letterfig5}
\centerline{\psfig{file=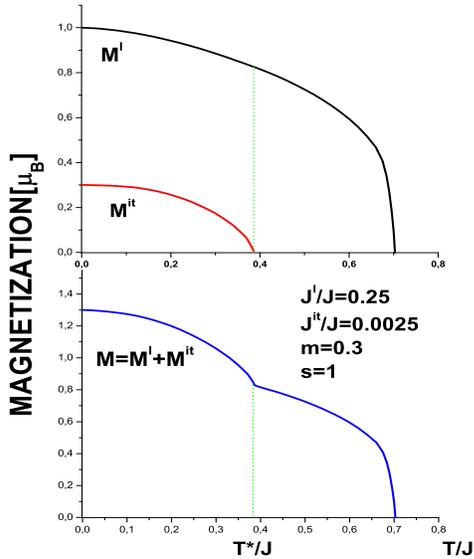,width=7cm,height=8cm}}
\caption{(color online)\,\, The magnetization of localized electrons $M^l(T/J)$-upper (black)line, the magnetization of the itinerant electrons $M^{it}(T/J)$-middle (red) line and the total magnetization $M(T/J)$-bottom (blue) line as a function of $T/J$ for parameters\, $s\,=\,1,\,m\,=\,0.3,\,J^l/J\,=\,0.25$ and $J^{it}/J\,=\,0.0025$. $T^*/J$- vertical dotted (green) line}\label{fig4}
\end{figure}
\end{center}

\section{\bf Theory and experiment}

The present paper is inspired from the experimental measurements of the magnetization-temperature curves
of $UGe_2$ \cite{2fmp5,2fmp6}. The existence of the characteristic temperature $T_x$ in the experimental measurements
and the present results (figure 4) refer us for assumption that the magnetic properties of $UGe_2$ are result of
two magnetic moments. One can write an effective Hamiltonian in terms of two vector fields
${\bf M}_{1i}$ and ${\bf M}_{2i}$ which identify the local orientation of the magnetizations (see Eq.\ref{letter14})
\begin{equation}
 h= -  J_1\sum\limits_{  \langle  ij  \rangle  } {{\bf M}_{1i} \cdot
{\bf M}_{1j}}-  J_2\sum\limits_{  \langle  ij  \rangle  } {{\bf M}_{2i}
\cdot {\bf M}_{2j}}
  - J\sum\limits_i {{\bf
M}_{1i}}\cdot {\bf M}_{2i}. \label{UGeH}
\ee
The exchange constants $J_1,\,J_2$ and $J$ are positive (ferromagnetic).

 Magnetism of $UGe_2$ is due to magnetic ordered moments of $5f$ uranium electrons.
They have dual character and in $UGe_2$ are more
itinerant than in many uranium compounds known as "heavy-fermion systems". The $LDA+U$ calculations show
the existence of well separated majority spin state with orbital projection $m_l=0$ \cite{Pickett1}.
This can be modeled with spin $1/2$ fermion. Then the local magnetization of the fermion ${\bf M}_{1i}$
is identical to ${\bf M}^{it}_i$ in the Hamiltonian Eq.\ref{letter14}. The saturation magnetization $m$ is close to $1/2$
at ambient pressure and decreases with increasing the pressure. The collective contribution of the others
uranium $5f$ electrons to the magnetization is described by ${\bf M}_{2i}$ vector with saturation magnetization $s=1$.
One thinks of these electrons as localized, but they are not perfectly localized in $UGe_2$. This means that
saturation magnetization $s$ could be smaller then one.

The $UGe_2$ compounds have strong magnetic anisotropy. It can be effectively accounted for introducing a gap in
the expressions for the dispersions Eq.(\ref{dispersion2}) . As a result the magnon of the system has a gap.
This is not important for the anomaly because it is a consequence of a different interactions
of the magnon with the transversal fluctuations of the magnetization vectors  
${\bf M}_{1i}$ and  ${\bf M}_{2i}$. This is
why the magnetic anisotropy is not accounted for. In this way I focus on the essential ingredients
which lead to the anomaly.

To proceed one uses the Holstein-Primakoff representation Eqs (\ref{rsw2}) for the  vectors
${\bf M}_{1i}$, ${\bf M}_{2i}$ and accomplishes the same calculations as in Section III. The obtained magnetization-temperature
curves, for different choices of model parameters, are depicted in figure 5. I set the Curie temperature
to be equal to the experimental one. This fixes the exchange constant $J$. The constants $J_1/J$ and $J_2/J$
are chosen so that the ratio $T_C/T^*$ to be close to the experimental value.
\begin{center}
\begin{figure}[htb]
\label{letterfig5}
\centerline{\psfig{file=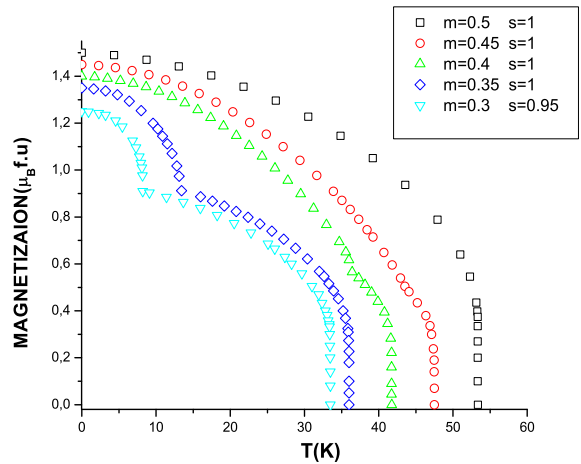,width=8.5cm,height=7cm}}
\caption{(color online)\,\,Magnetization-temperature curve obtained within an effective two
magnetic ordered moments model of $UGe_2$ magnetism}
\end{figure}
\end{center}

The first curve from above (black squares) is calculated for parameters $m=0.5,\,s=1,\,J_1/J=0.0005$ and $J_2/J=0.05$.
The strong interaction between itinerant and "localized" electrons aligns their magnetic orders
so strong  that they become zero at one and just the same temperature $T_C$. The magnetization-temperature curve
is typical Curie-Weiss curve. The result is different if the exchange constant $J$ is relatively smaller.
The ferromagnetic phase is divided into two phases: low temperature phase $0<T<T^*$ where all $5f$ uranium electrons
give contribution to the magnetization, and high temperature ferromagnetic phase $T^*<T<T_C$
where the contribution to the magnetization of itinerant electrons is zero.
The next curve (red circles) is obtained for parameters $m=0.45,\,s=1,\,J_1/J=0.0016,\,J_2/J=0.16$,
the third one (green triangles) for parameters  $m=0.4,\,s=1,\,J_1/J=0.0018$ and $J_2/J=0.18$,
the fourth curve (blue rhombs) corresponds to parameters  $m=0.35,\,s=1,\,J_1/J=0.004$ and $J_2/J=0.4$,
and for the last one $m=0.3,\,s=0.95,\,J_1/J=0.0057$ and $J_2/J=0.57$.
The curves show that increasing the constants $J_1/J$ and $J_2/J$ the ration $T_C/T^*$ increases ($T_C/T^*=1,\,1.092,\,1.46,\,2.68
,\,4.08$), and
$T^*$ approaches to zero ($T^*=53.35K,\,43.511K,\,36.433K,\,13.44K,\,8.21K$).
Comparing with experiments \cite{2fmp3,2fmp6} one concludes that increasing the pressure
the exchange constant between itinerant and "localized" electrons $J$ increases, but
exchange constants between itinerant electrons $J_1$ and between localized electrons $J_2$ increase faster,
so that the ratios  $J_1/J$ and $J_2/J$ increase.

The anomalous temperature dependence
of the ordered moment, known from the experiments with $UGe_2$
\cite{2fmp2,2fmp5,2fmp6,2fmp7}, is very well reproduced theoretically in the present paper (figure 5).
Below $T_x$ ($T^*$ in the present paper) the ferromagnetic
moment increases in an anomalous way. The low temperature, large moment phase is referred to as $FM2$,
while the high temperature low-momenta phase is referred to as $FM1$ \cite{2fmp7,Pfleiderer2}.
The present theoretical result gives new insight into $FM1\rightarrow FM2$ transition. It is shown that
between Curie temperature and $T^*<T_C$ the contribution of the itinerant $UGe_2$ electrons to the magnetization is zero.
They start to form magnetic moment at $T^*$.

There are experiments which support the present theoretical result. The measurements \cite{2fmp5}
show that the resistivity display a down-turn around $T^*(=T_x)$, that is best seen in terms of a broad
maximum in the derivative $d\rho/dT$ \cite{Oomi1}. It is well known that the onset of magnetism in the itinerant
systems is accompanied with strong anomaly in resistivity \cite{2fmp12}. The experiments  \cite{2fmp5,Oomi1} prove that
only part of $5f$ uranium electrons start to form magnetic order at Curie temperature. The other ones do this
at temperature $T_x(=T^*)$ well below $T_C$, in agreement with the theoretical result.
Further evidence for the nature of the
$FM1\rightarrow FM2$ transition has been observed in the high resolution photoemission, which show
the presence of a narrow peak in the density of states below $E_F$ that suggests itinerant
ferromagnetism \cite{Ito}.

\section{\bf Summary}
In summary, it is obtained an effective theory of two magnetic ordered vectors from spin-Fermion model.
I have worked out a renormalized spin-wave theory and its extension to describe
the two ferromagnetic phases of a spin-Fermion system: high temperature phase $T^{*}<T<T_C$,
where only localized spins form magnetic moment, and low temperature
phase  $0<T<T^{*}$, where localized spins and itinerant electrons contribute the ordered ferromagnetic
moment.

It is important to stress that the two ferromagnetic phases can not be obtained within RKKY theory because it
utilizes only the transversal fluctuations of the localized spins. Integrated over the spinless Fermions we obtain
the exchange interaction between transversal fluctuations of the localized and itinerant spins instead of RKKY exchange.
This point is basic for
the understanding of the two ferromagnetic phases in the spin-Fermion systems.

The present theory of magnetism permits to consider more complicated systems such as the $UGe_2$ compound.
The effective model, in terms of two magnetic ordered moments, reproduces very well the experimental
magnetization-temperature curves. The results give new understanding of the two ferromagnetic phases.
The large moment phase ($FM2$) is a phase where all $5f$ uranium electrons contribute the magnetization,
while the electrons are partially ordered in the low-momenta phase ($FM1$). The result differs from scenarios
studied in the literature \cite{2fmp8a,Lonzarich03,Pickett04}, and is important for the study of the coexistence
of ferromagnetism and superconductivity in these compounds.

\section{Acknowledgments}
This work was partly supported by a Grant-in-Aid DO02-264/18.12.08 from NSF-Bulgaria. 
The author acknowledges the financial support of the Sofia
University under Grant No. 051/2010.

\begin{appendix}
\section{}

To make more transparent the derivation of the Hamiltonian in the Hartree-Fock approximation Eq.(\ref{HF1})
I consider the first term in the Hamiltonian of the magnon-magnon interaction Eq.(\ref{heff2}).
To write this term in the Hartree-Fock  approximation one represents the product of two Bose operators in the form
\be\label{App1}
a^+_i a_j\,=\,a^+_i a_j\,-\,<a^+_i a_j>\,+\,<a^+_i a_j> \ee and neglects all terms $(a^+_i a_j\,-\,<a^+_i a_j>)^2$ in the four magnon interaction Hamiltonian. The result is
\bea\label{App2}
\frac 12 a^+_i a_j a^+_i a_i&\approx&-<a^+_i a_{j}><a^+_i a_i> \nonumber \\& + & <a^+_i a_{j}> a^+_i a_i+a^+_ia_{j}<a^+_i a_i> \nonumber \\
\frac 12 a^+_{j} a_i a^+_{j} a_{j} &\approx & -<a^+_{j} a_i><a^+_{j} a_{j}> \nonumber \\ & + & <a^+_{j} a_i> a^+_{j} a_{j}+a^+_{j} a_i <a^+_{j} a_{j}> \nonumber \\
\frac 12 a^+_{j} a_j a^+_i a_{j} &\approx & -<a^+_{j} a_{j}><a^+_i a_{j}>  \\ & + & <a^+_{j} a_{j}>a^+_i a_{j}+a^+_{j} a_{j}<a^+_r a_{j}> \nonumber\\
a^+_i a_i a^+_{j} a_{j} &\approx & -<a^+_i a_i>< a^+_{j} a_{j}> \nonumber \\ & + & <a^+_i a_i> a^+_{j} a_{j}+a^+_i a_i <a^+_{j} a_{j}> \nonumber \\
&-& <a^+_i a_{j}>< a^+_{j} a_i> \nonumber \\ & + &  <a^+ _i a_{j}> a^+_{j} a_i+a^+_{j} a_i <a^+_i a_{j}> \nonumber
\eea

The Hartree-Fock approximation of this part of the Hamiltonian of magnon-magnon interaction reads
\bea\label{App3}
& & \frac 14 J^l \sum\limits_{< ij>}\left[a^+_i a^+_j( a_i-a_j)^2 + (a^+_i- a^+_j)^2  a_i a_j\right] \nonumber\\  & \approx & 3NJ^l s^2 \left(u^l-1\right)^2 \\
& + & s J^l \left(u^l-1\right)\sum\limits_{< ij>}\left( a^+_i a_i\,+\,a^+_{j} a_{j}\,-\,a^+_{j} a_i\,-\,a^+_i a_{j}\right) \nonumber \eea
where the Hartree-Fock parameter $u^l$ is defined by the equation
\be\label{App4}
u^l\,=\,1\,-\,\frac {1}{3 s}\frac 1 N \sum\limits_{k}\varepsilon_k <a^+_k a_k> \ee
Combining the $a$-bosons' part of the Hamiltonian Eq.(\ref{letter17}) (the first term) and  Eq.(\ref{App3}) one obtaines the Hartree-Fock approximation for the $a$-bosons' part of the Hamiltonian Eqs.(\ref{HF2},\ref{HF3}).
\bea\label{App5}
H^a & \approx & 3NJ^l s^2 \left(u^l-1\right)^2 \\
& + & s J^l u^l \sum\limits_{< ij> }\left( a^+_i a_i\,+\,a^+_{j} a_{j}\,-\,a^+_{j} a_i\,-\,a^+_i a_{j}\right) \nonumber \eea
In the same way one obtains the Hartree-Fock approximation of the $b$-bosons'  and inter bosons' parts of the Hamiltonian. 
The result is the $h_{HF}$ Hamiltonian Eqs.(\ref{HF2},\ref{HF3}).

To calculate the thermal average $ <a^+_k a_k>$, in the Eq.(\ref{App4}), one utilizes the Hamiltonian $h_{HF}$. 
Therefor, the matrix element  depends on the Hartree-Fock parameters, and equation (\ref{App4}) is one of  
the self consistent equations for these parameters.

The matrix element can be represented in terms of $\alpha_k (\alpha_k^+)$ and $\beta_k (\beta_k^+)$ by means 
of equations(\ref{letter21})
\be\label{App6}
<a^+_k a_k>\,=\,\cos^2\theta_k \,n_k^{\alpha}\, +\, \sin^2 \theta_k\, n_k^{\beta} \ee
where $n_k^{\alpha}=<\alpha_k^+\alpha_k>,\,n_k^{\beta}=<\beta_k^+\beta_k>$ are the Bose functions of $\alpha$ and $\beta$ excitations.
Substituting the thermal average in Eq.(\ref{App4}) with Eq.(\ref{App6}), one obtains that equation (\ref{App4}) 
is exactly the first equation of the system Eq.(\ref{rsw14}) which in turn is obtained from the first of the equations (\ref{rsw13}).

\end{appendix}

\end{document}